\documentclass{article}
\usepackage{latexsym,epsfig}

\begin{document}
A model of underground ridership during the severe outbreaks of the SARS 
epidemic in a modern city

Kuo-Ying Wang

Department of Atmospheric Sciences

National Central University

Chung-Li, Taiwan

\textbf{Abstract }The outbreaks of the severe acute respiratory syndrome 
(SARS) epidemic in 2003 resulted in unprecedented impacts on people's daily 
life. One of the most significant impacts to people is the risk of 
contacting SARS while engaging daily routine activity. In this work we use 
data from daily underground ridership in Taipei and daily reported SARS 
cases in Taiwan to model the dynamics of the public underground usage during 
the wax and wane of the SARS period. We found that for each reported SARS 
case there is an immediate loss of about 1200 underground ridership. These 
loss rates propagate to the following days with an $e$-folding decay time of 
about 28 days, reflecting the public perception on the risk of contacting 
SARS disease when travelling with the underground system. About 50{\%} of 
daily ridership was lost during the peak of the 2003 SARS period, compared 
with the loss of 80{\%} daily ridership during the closure of the 
underground system after Typhoon Nari, the loss of 50-70{\%} ridership due 
to the closure of the governmental offices and schools during typhoon 
periods, and the loss of 60{\%} daily ridership during Chinese New Year 
holidays. 


\textbf{1. Introduction}

The 2003 SARS epidemic is a recent vivid example, demonstrating the deep 
impacts that an infectious disease can have on human society. For example, 
TIME magazine called Taiwan a SARS Island (16), that SARS sinks Taiwan (17), 
and China as a SARS Nation (18). One of the best ways to understand the 
response of people living through an emerging disease is by examining the 
change of people's daily activity with respect to the variations of the 
reported SARS cases during the epidemics. However, there is a lack of study 
showing the dynamics of public's response to the perceived risk associated 
with the daily reported SARS cases. Previous works studied the dynamics of 
the daily accumulated infected cases during the SARS outbreaks in Beijing, 
Canada, Hong Kong, Singapore, and Taiwan (1, 2, 3, 12), respectively. A 
study examined the impact of influenza on death rates in tropical Singapore 
(4). Since confined environment with short distance is conducive for 
infectious transmission between species (5, 6), the perceived risk 
associated with staying in any confined space during the SARS epidemics 
alters people's daily activity. In this work, we use the Taipei underground 
mass transportation system, which is a typical of confined space, and 
reported SARS cases in Taiwan, to show that the dynamics of the underground 
usage can be modeled by the daily variations of the reported SARS cases.

\textbf{2. Methods}

\textbf{2.1. Daily Underground Ridership Data}

The Taipei underground system transports about 1 million people per day (7). 
These daily ridership exhibits a strong weekly cycle. A lower amount of 
people traveling on Wednesday (a short weekend), a weekly peak on Friday 
(before the long weekend), the lowest amount of people traveling on Saturday 
and Sunday, and the rest of the week days are about the same. Except for 
occasional events such as typhoon and the Chinese New Year (8), the weekly 
pattern is roughly the same through the year. This stability in the daily 
ridership provides a good quantifiable measure of public reaction when an 
unprecedented risk occurs in the society.

Since weekly patterns of the passengers are less perturbed during the weeks 
from early spring to early summer than other periods, we can determine mean 
daily underground ridership $\overline p $in a week based on the average of 
the twelve weeks, started from the week with the first Monday in March, for 
the years 2001, 2002, 2004, and 2005, respectively. The statistical daily 
ridership $\overline p $ for 2003 is calculated as the mean of the daily 
ridership from 2001, 2002, 2004, and 2005. Hence, if without the disturbance 
of significant factors such as an approaching typhoon, a long holiday, 
festivals, and epidemics, the daily ridership will normally maintain a 
constant pattern throughout the week. 

\textbf{2.2. A Dynamic Model }

In order to model the daily variations of the underground ridership with 
respect to the daily reported SARS cases, a dynamical model was developed to 
simulate day-to-day variations of the underground ridership in the periods 
before, during, and after the SARS epidemics. Since the model is dynamic in 
nature, the model variables and the external forcing that governs the time 
evolution of the model variables must be established so that the model is 
able to make prediction based on the change of external forcing. 

During the 2003 SARS period, we observed two significant relationships 
between the daily underground ridership and the daily reported SARS case. 
Firstly, there exists a quick response of the underground ridership with 
respect to the daily reported SARS cases that made headlines almost everyday 
in the mass media during the SARS period. The overwhelming reports from 
these public media appear to have big impacts on the willingness of the 
public in using the underground as a mean for going to schools and offices 
(both schools and offices were not closed during the SARS period). The 
public perception of risk in associating with the use of the underground 
system vividly reflected in the significant drops of underground ridership 
(9). Secondly, the gradual increases in underground ridership during the 
final stage of the SARS epidemic. This indicates the return of public 
confidence in using the underground system as a mean for daily 
transportation to offices and schools. These two observations indicate that 
a dynamic model should represent these effects, i.e., the sharp drops in the 
ridership associated with increases in the reported SARS cases, and a 
gradual return of the underground ridership as the reported SARS cases 
gradually faded away from the headlines. 

Based on the daily ridership from 2001 to 2005 and daily reported SARS cases 
in 2003, we can write a model of daily ridership with respect to the daily 
reported SARS cases as
\[
p(t)=\overline p -\sum\limits_{i=1}^{i=t} {L(i)\exp } (-(t-i)/\tau )
\]
Here, on a given day $t$, $\overline p $is the daily normal underground usage 
predicted by the statistics as described above;$L=kc(t)$ is the loss of the 
underground ridership, due to the perceived risk if traveling with the 
underground system; $k$ is the loss rates of the underground ridership per 
reported SARS case; $c$ is the daily reported SARS cases; $i$ is the day number 
since the simulation started (1 Mar 2003); and$\tau $is the $e$-folding time. 

Notice that the real-time reported probable SARS cases (9) were used in this 
work. These were the information affecting people's decision during the 
height of the SARS epidemics. The SARS cases published after the SARS 
epidemics are slightly different (10, 11, 12). The instantaneous passenger 
loss rate on day $i$ will propagate exponentially to the following days with an 
$e$-folding time of $\tau $days. Here the e-folding time measures the damping 
of the perceived risk factor due to each new reported SARS cases, reflecting 
the public expectation of the risk of contacting the disease arises from 
each newly reported SARS cases. Hence, the amount of passengers at each day 
$t$ is determined by the daily normal ridership$\overline p $, subtracted the 
loss of ridership due to the number of reported SARS cases at that day and 
the accumulated impacts from the days before. 

\textbf{3. Results}

\textbf{3.1. The Underground Ridership During Non-SARS Years}

\begin{figure}[htbp]
\centerline{\includegraphics[width=5.88in,height=6.01in]{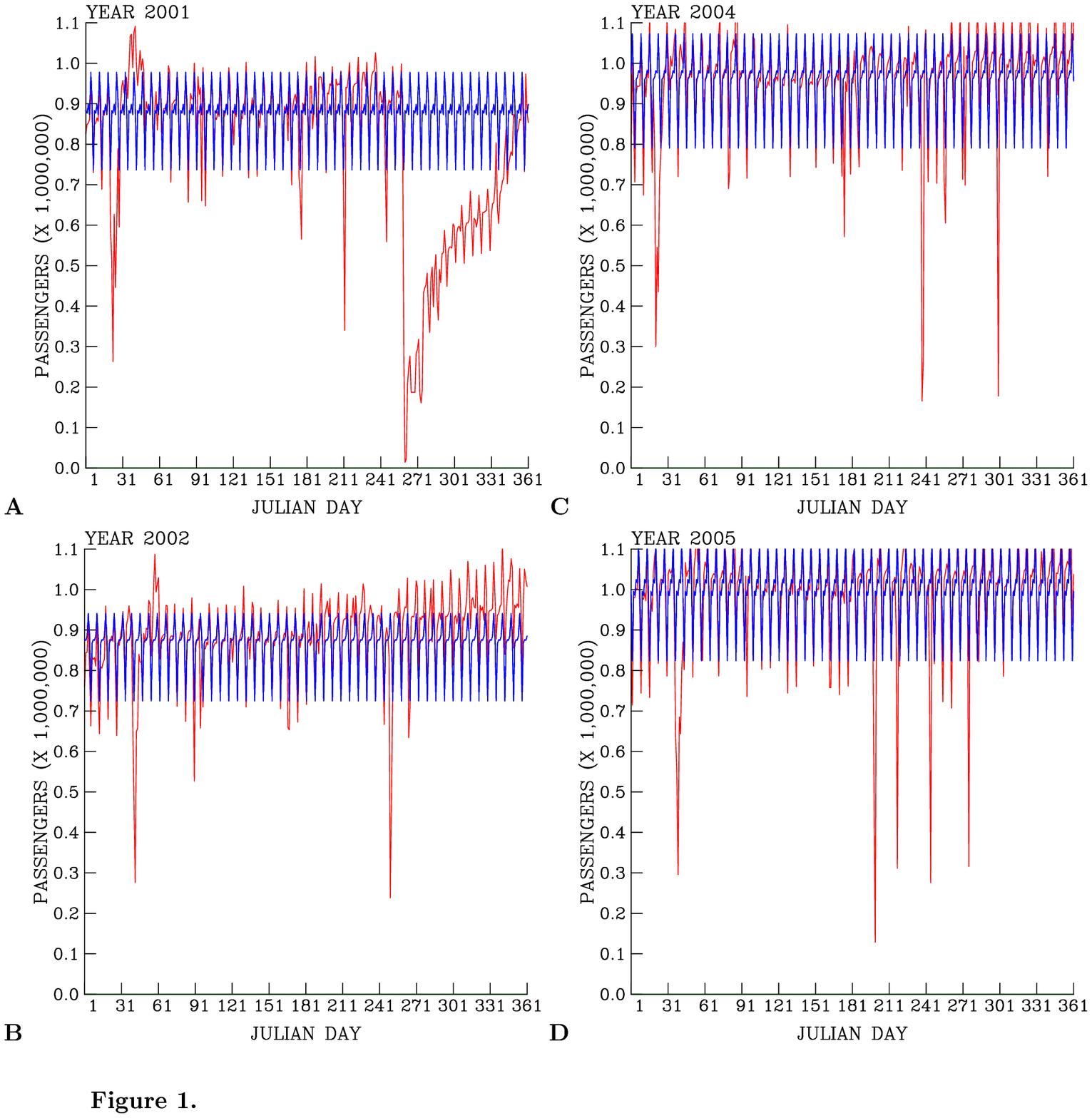}}
\caption{\label{fig2} Daily underground passengers (red curve), and the passengers predicted by
the statistical model (blue line) in 2001 (\textbf{A}), 2002 (\textbf{B}),
2004 (\textbf{C}), and 2005 (\textbf{D}).}
\end{figure}

Figure 1 shows a time-series plot of recorded and modeled daily underground 
ridership during the non-SARS years of 2001, 2002, 2004, and 2005, 
respectively. During these years, the daily underground ridership can be 
approximated by the statistical average daily ridership ($\overline p )$ of 
each year. During most of the days the actual amount of people traveling 
with the underground is close to the statistical predictions, indicating 
that the underground usage during normal days are very regular. However, 
there are days when the model and the actual underground ridership show big 
discrepancies. These are the days when special events (e.g., Chinese New 
Year, spring holiday, and typhoons) occurred. For example, Figure 1B shows a 
typical example of the time-series plot of daily ridership using the Taipei 
underground system in 2002. We note that the first big drop in ridership 
after day 31 corresponds to the Chinese New Year holiday, the second big 
drop in numbers after day 91 is due to students' spring holiday, and the 
drop after day 241 Julian is due to a typhoon. Similar situations applied to 
other years as well. We note that a big drop of ridership during the days 
241-271 in 2001 is due to the closure of two main lines of the underground 
system, which was flooded by the severe rainfall during the passing of 
Typhoon Nari (8).

\begin{figure}[htbp]
\centerline{\includegraphics[width=5.88in,height=6.01in]{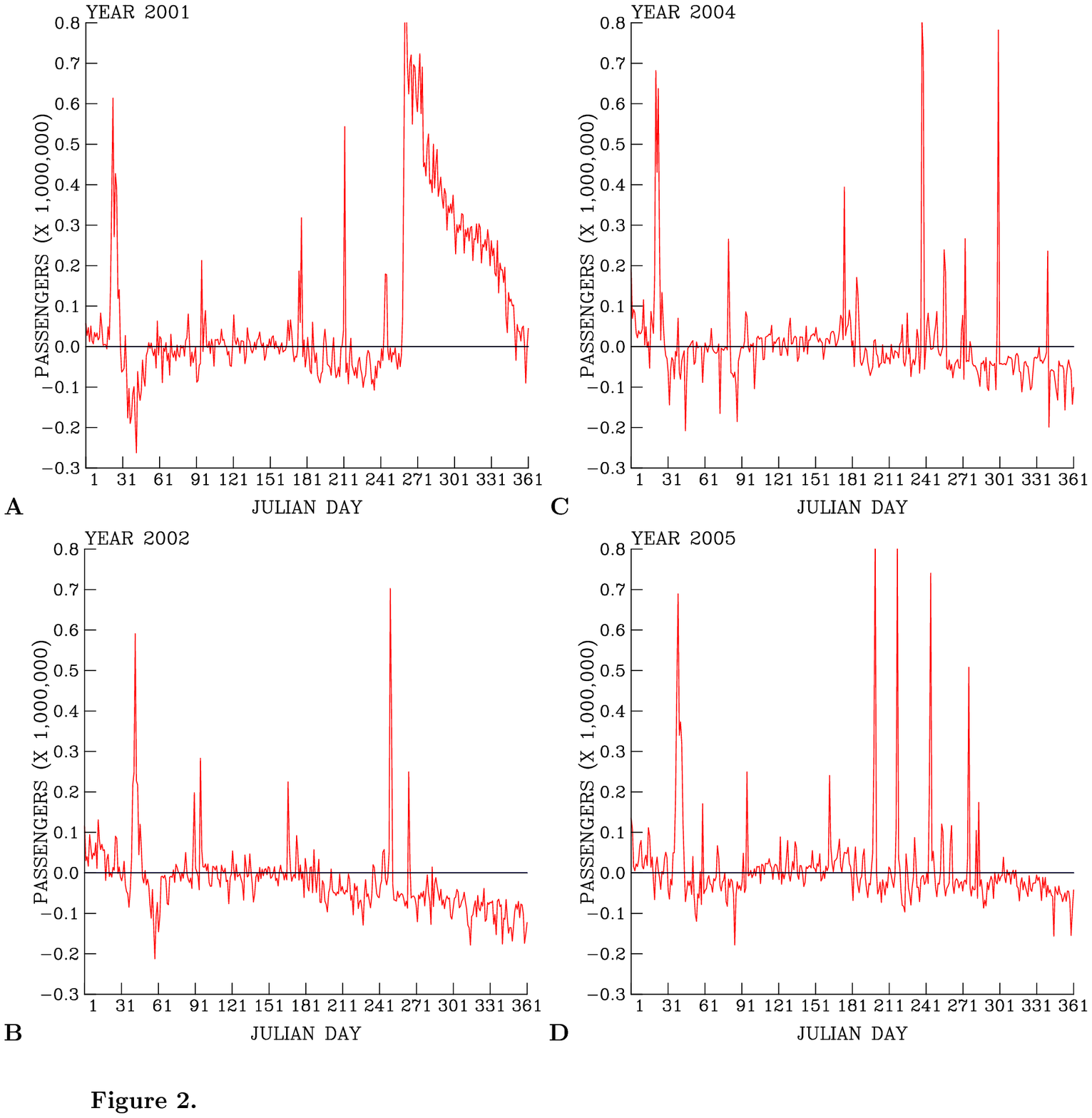}}
\caption{\label{fig2} Difference in daily ridership between the model prediction and the actual
daily ridership in 2001 (\textbf{A}), 2002 (\textbf{B}), 2004 (\textbf{C}),
and 2005 (\textbf{D}).}
\end{figure}

Figure 2 further compares the discrepancies in ridership between the 
statistical predictions and the actual numbers. During the spring months 
(days 61-150) of each year, the statistical model predicts the daily 
ridership that are, in most of the days, close to within 10{\%} of the 
actual numbers of people taking the underground. Exceptions occur during 
rare events (e.g., long holidays, festivals, typhoons, etc). We find that 
the underground flooding in 2001, casued by Typhoon Nari, resulted in a loss 
of about 80{\%} daily ridership; the Chinese New Year holiday causes a loss 
of 60-70{\%} daily ridership; and the close of the governmental offices and 
schools during typhoon periods caused a loss of 50-80{\%} daily ridership. 
The increases in people taking the underground toward the end of 2002, 2004, 
and 2005, respectively, could be due to the factors that the underground 
ridership was steadily growing and also more people tend to take underground 
during the winter months. 

\textbf{3.2. The Underground Ridership During the 2003 SARS Year}

\begin{figure}[htbp]
\centerline{\includegraphics[width=5.88in,height=6.01in]{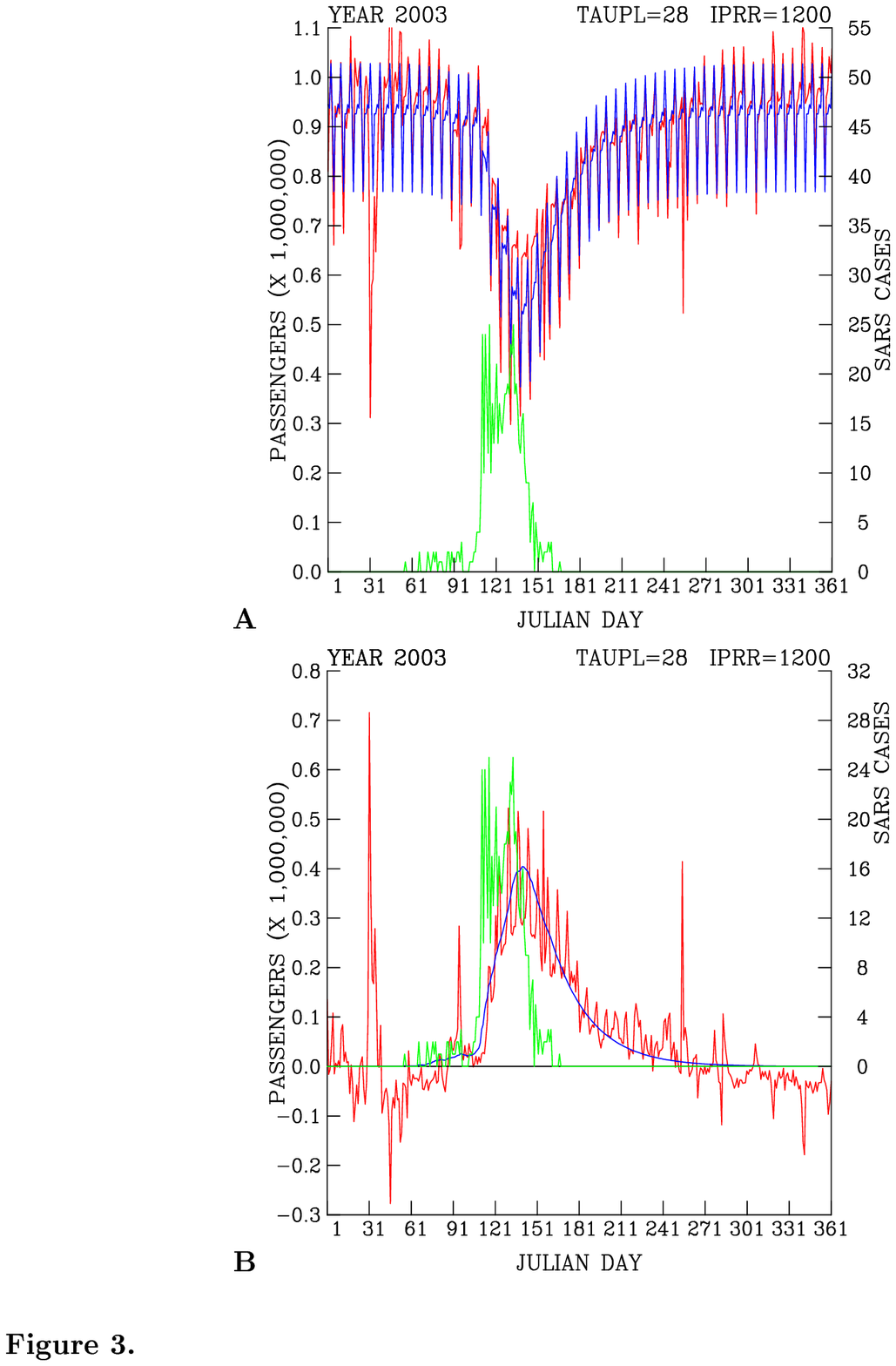}}
\caption{\label{fig3} Predicted (blue curve) and actual (red curve) daily ridership (\textbf{A}).
Difference in the actual daily ridership (red curve) and the predicted
ridership (blue curve) (\textbf{B}).Green curve shows daily reported SARS
cases. Both data are for 2003.}
\end{figure}

In a sharp contrast with the normal underground usage during 2001, 2002, 
2004, and 2005, the daily ridership in 2003 shows anomalously high loss of 
ridership from about day 60 to days 120-150, when the maximum reduction of 
daily ridership of half a million were occurred (Figure 3). About 50{\%} of 
daily ridership was lost during peak of the 2003 SARS periods. This period 
concurs with the SARS outbreaks in Taiwan (10, 11). The reason for the drop 
of daily underground passengers is clearly related to the rising numbers of 
reported probable SARS cases (9) during this period. Figure 3A compares the 
time evolution of SARS cases and the wane and the wax of the daily 
underground ridership. The peak of the reduction in the daily ridership 
occurred after the peak of the reported probable SARS cases. While the 
reported SARS cases drop sharply during days 151-181, the returns of the 
ridership to the underground appear to be at a slow pace during days 
151-271. Predicted loss of the daily underground ridership and its 
comparison with the actual ridership are shown in Figure 3B. The sharp 
response in daily ridership following the increase of the reported SARS 
cases, and the slow return of the ridership after the peak of the SARS cases 
is well reproduced by the model, Figure 3B. The close agreement between the 
model and actual underground ridership indicate that the model can 
successfully reproduce the daily underground ridership during the 2003 SARS 
epidemics in Taiwan. Though the fear of the SARS still linger on in 2004, no 
reported SARS cases in that year resulted in the normal use of the 
underground system as seen from the model and the actual daily ridership, 
Figure 2C.

\textbf{3.3. Sensitivity of Underground Ridership to Reported SARS Cases}

\begin{figure}[htbp]
\centerline{\includegraphics[width=5.88in,height=6.01in]{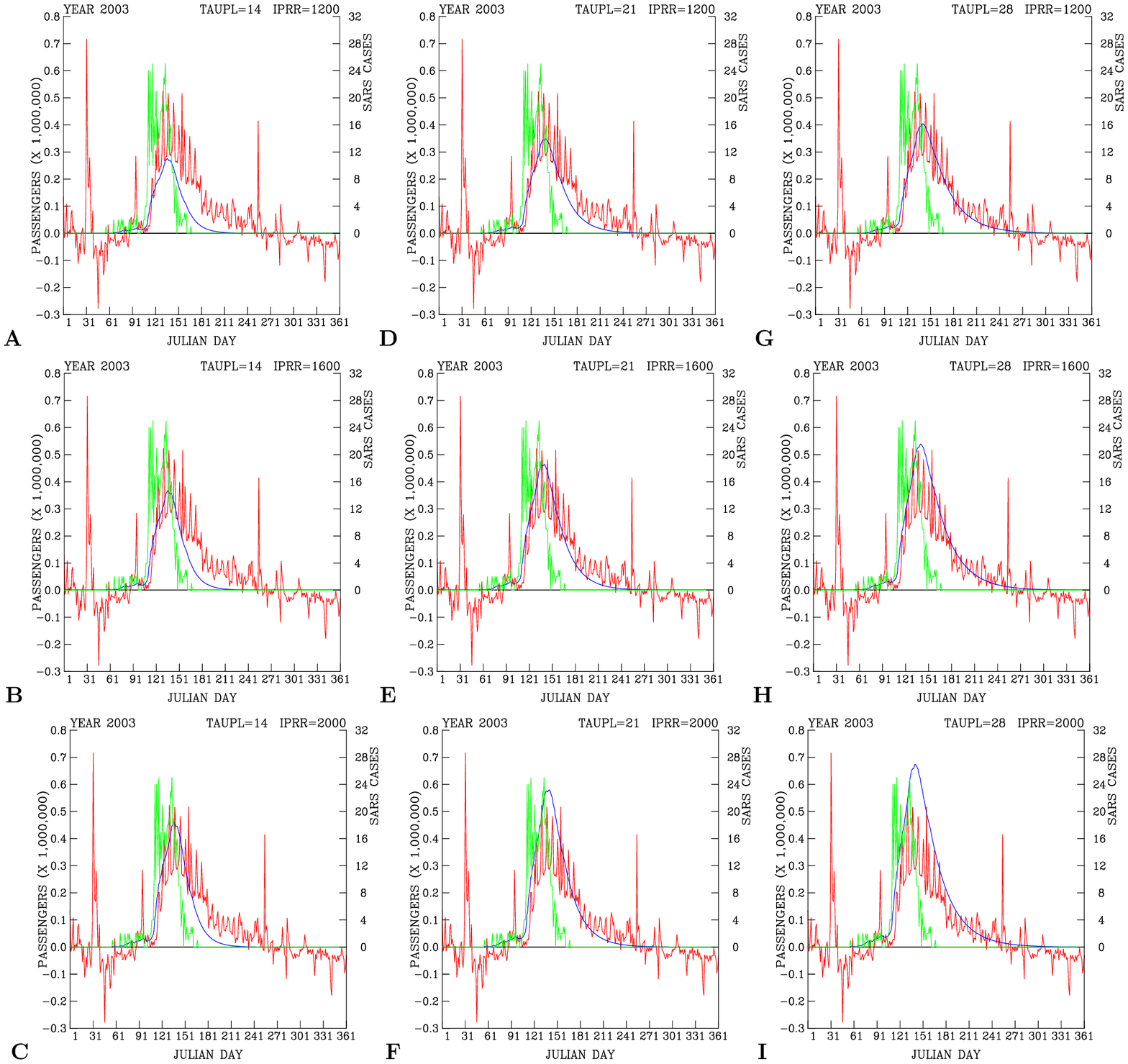}}
\caption{\label{fig4} Difference in the actual daily ridership (red curves) and the predicted
daily ridership (blue curves) with respect to a combination of three
instantaneous passenger reduction rates ($k)$ and three passenger $e$-folding time
($\tau )$ scales. For plots in the columns from the left to the right
showing $\tau $=14, 21, and 28 days, respectively. For plots in the rows
from the bottom to the top showing $k$=1200, 1600, and 2000 ridership loss
rates per reported SARS case, respectively. Green curves show daily reported
SARS cases. These data are shown for 2003.}
\end{figure}

Two parameters are keys to the predicted underground ridership with respect 
to the daily reported SARS cases: Instantaneous ridership loss rate ($k)$ per 
reported SARS case, and the $e$-folding time ($\tau )$ indicating the 
propagation of the loss rates to subsequent days. Figure 4 shows tests of 
various values of these two parameters. For the same $e$-folding time (the 
periods that perceived risk lasts), for example $\tau $=14 days (Figures 
4A-C), the larger the daily ridership loss rates $k$ per reported SARS case 
(degree of shocks to the public), the deeper the reduction in the 
underground ridership will be resulted. But the time to return to the normal 
daily ridership is similar for different ridership loss rates after passing 
the peaks in the SARS cases. These results indicate that, if the time scales 
of public perception to each reported SARS case are the same, then the 
impact on the loss of underground passengers will be limited to the days 
close to the peak of the reported SARS cases. 

On the other hand, if the passenger loss rates are the same, for example 
$k$=1200 (Figures 4A, 4D, and 4G), then the longer the $e$-folding time scale 
$\tau $, the slower the return of the underground ridership to the normal. A 
long $e$-folding time scale also results in a large accumulated loss due to the 
accumulated effects from previous days (Figure 4G). Hence, long period of 
the public perception of the risk associated with the reported SARS cases is 
likely to cause the long-lasting impact on the behavior of people and their 
willingness to use the underground. 

\textbf{4. Summary}

In this work we show that the dynamics of the Taipei underground usage 
during the 2003 SARS epidemic in Taiwan are closely linked to the daily wax 
and wane of the reported probable SARS cases. Our model shows that each 
reported SARS case results in an immediate loss of about 1200 underground 
ridership, reflecting the public perception of immediate risk associated 
with the intense report of the SARS outbreaks and their reluctance in using 
the underground system. The public perception of the risk propagates and 
exponentially decays to the following days with an $e$-folding time of about 28 
days. This duration of time reflects the perception of the risk perceived by 
the normal underground passengers. Our study shows that the longer the 
$e$-folding time (perception of the risk), the slower the return of the 
underground ridership. A huge loss of the underground ridership but with a 
short $e$-folding time results in predicted passengers returning to the 
underground system sooner than what had occurred. The combination of the 
immediate passenger loss rates and their impacts propagates to the following 
days resulting in the occurrence of the peak of the ridership loss later 
than the peak of the reported SARS cases. About 50{\%} of daily ridership 
was lost during the peak of the 2003 SARS periods, compared with the loss of 
80{\%} daily ridership during the closure of the underground system after 
Typhoon Nari, the loss of 50-70{\%} ridership due to the closure of the 
governmental offices and schools during typhoon periods, and the loss of 
60{\%} daily ridership during Chinese New Year holidays.

Since social distancing measures have been shown to be important for 
containing an emerging disease (6, 13, 14, 15, 22), our results could be 
useful in incorporating into the disease spreading models where underground 
usage is an important connection node for social behaviors. There are other 
major cities such as Hong Kong, Singapore, and Beijing all contain massive 
underground systems and were impacted by the 2003 epidemics (19, 20, 21). 
Our model developed here could be useful to test if similar ridership 
behaviors found in Taipei are also applicable to these major cities in Asia. 
In the context of avian flu, the underground ridership occurred under the 
SARS epidemics may provides us a glimpse on what the general public will 
response in the wake of next epidemics.

\textbf{Acknowledgement}

The author dedicates this work to those who suffered the SARS disease; the 
US CDC who helped Taiwan fights the SRAS war; and the doctors, nurses, 
voluntary workers, and public officers who stayed on duty during the 2003 
SARS epidemic. The author thanks P. Hadjinicolaou, O. Wild, A. Polli, and 
H.-C. Lee for their comments on the manuscript.

\textbf{References}

\begin{enumerate}
\item Hsieh YH, Cheng YS, Real-time forecast of multiphase outbreak, Emerg Infect Dis. 2006; 12:114-121.
\item Cauchemez S, Bo\"{e}lle PY, Donnelly CA, Ferguson NM, Thomas G, Leung GM, Hedley AJ, Anderson RM, Valleron AJ, Real-time estimates in early detection of SARS, Emerg Infect Dis. 2006; 12:110-113.
\item Zhou G, and Yan G, Severe Acute Respiratory Syndrome epidemic in Asia, Emerg Infect Dis. 2003; 9:1608-1610.
\item Chow A, Ma S, Ling AE, Chew SK, Influenza-associated deaths in tropical Singapore, Emerg Infect Dis. 2006; 12:114-121.
\item Lowen AC, Mubareka S, Tumpey, TM, Garc\'{\i}a-Sastre A, Palese P, The guinea pig as a transmission model for human influenza viruses, PNAS 2006; 103:9988-9992.
\item Ferguson NM, Cummings DAT, Cauchemez S, Fraser C, Riley S, Meeyai A, Iamsirithaworn S, Burke DS, Strategies for containing an emerging influenza pandemic in southeast Asia, Nature 2005; 437:209-214.
\item Taipei Rapid Transport Corporation (TRTC). Available at \underline {http://www.trtc.com.tw}. Statistics for daily passengers were published on-line since March 1996.
\item Wang KY, Shallcross DE, Hadjinicolaou P, Giannakopoulos C, Ambient vehicular pollutants in the urban area of Taipei: Comparing normal with anomalous vehicle emissions, Water Air Soil Pollu. 2004; 156:29-55.
\item Department of Health, Taipei City Government. Available at \underline {http://sars.health.gov.tw/INDEX.ASP}. The probable SARS cases during the SARS outbreak were published and updated real-time at \underline {http://sars.health.gov.tw/article.asp?channelid=C{\&}serial=262{\&}click=}.
\item Lee ML, et al., Severe acute respiratory syndrom -- Taiwan, 2003, MMWR Morb. Mortal Wkly Rep. 2003; 52\textbf{: }461-466.
\item Chen YC, et al., SARS in hospital emergency room, Emerg Infect Dis. 2004; 10:782-788.
\item Hsieh YH, Chen CWS, Hsu SB, SARS outbreak, Taiwan, 2003, Emerg Infect Dis. 2004; 10:201-206.
\item Fraser C, Riley S, Anderson RM, Ferguson NM, Factors that make an infectious disease outbreak controllable, PNAS 2004; 101:6146-6151.
\item Germann TC, Kadau K, Longini IM Jr., Macken CA, Mitigation strategies for pandemic influenza in the United States, PNAS 2006; 103:5935-5940.
\item Longini IM Jr., Nizam A, Xu S, Ungchusak K, Hanshaoworakul W, Cummings DA, Halloran ME, Containing pandemic influenza at the source, Science 2005; 309:1083-1087. 
\item Time, 26 May 2003, page 3.
\item Time, 2 June 2003, page 3.
\item Time, 5 May 2003. Cover story.
\item Donnelly CA, et al., Epidemiological determinants of spread of causal agent of severe acute respiratory syndrome in Hong Kong, Lancet 2003; 361:1761-1766.
\item Zhou G, Yan G, Severe acute respiratory syndrome epidemic in Asia, Emerg Infect Disc. 2003; 9:1608-1610.
\item Dye C, Gay N, Modeling the SARS epidemic, Science 2003; 300:1884-1885.
\item Eubank S, Guclu H, Anil Kumar VS, Marathe MV, Srinivasan A, Toroczkai Z, Wang N, Modelling disease outbreaks in realistic urban social networks, Nature 2004; 429:180-184.
\end{enumerate}

\end{document}